\documentclass[10pt,letterpaper]{article}
\usepackage{opex3}
\usepackage{subfigure}
\usepackage{cite}

\begin{document}

\title{Binocular parallax stereo imaging based on correlation matching algorithm}

\author{Yongchao Zhu, Hu Li, Jianhong Shi*, Ying Yang, \\Fei Cao, and Guihua Zeng}

\address{State Key Laboratory of Advanced Optical Communication Systems and Networks,
Department of Electronic Engineering, Shanghai Jiao Tong University, Shanghai
200240, P.R.China. }

\email{*purewater@sjtu.edu.cn} 

\begin{abstract}
The intensity fluctuation correlation of pseudo-thermal light can be utilized to realize binocular parallax stereo imaging (BPSI). With the help of correlation matching algorithm, the matching precision of feature points can reach one pixel authentically. The implementations of the proposed BPSI system with real objects were demonstrated in detail. And the experimental results indicated that the proposed system performs better when the object's superficial characteristics are not obvious, for example its surface reflectivity is constant.
\end{abstract}



\section{Introduction}
Multifarious three dimensional sensing technologies have been investigated for many decades. Since along with rapid technological progress, more and more 3D imaging application scenarios turn up, the research on 3D imaging needs to continue. Binocular parallax stereo imaging or multi-view stereo imaging \cite{Burckhardt1968,Yang1988,Son2005,Xiao2013,Javidi2009,Rubio2011,Sung-Keun2013} belongs to one of early stage three dimensional imaging technologies. Making use of the parallaxes of multi-view images and the depth sensing theory, the 3D reconstruction map can be calculated according to geometry. The depth resolution in BPSI depends on the binocular distance and the matching precision of feature points. Normally in traditional matching algorithm of BPSI, the feature points are extracted by searching distinctive distributions. Therefore there exist matching errors in traditional BPSI. Moreover, not all points of the object are feature points. Then if the object's images have a weak variation in distribution, the 3D reconstruction is likely to be distorted.

Recently, a 3D computational imaging with single-pixel detectors \cite{Sun2013} was proposed in the journal of Science, in which computational ghost imaging with a digital micromirror device (DMD) was introduced to combine with the multi-view imaging. The great advantages of this method are the capabilities of realizing 3D imaging without a spatially resolving detector and using a broadband light source. However this method needs a long time capturing and has the similar deficiency with the traditional BPSI as well.

Ghost imaging or correlation imaging \cite{Pittman1995,Boyd2002,Gatti2004,Wu2009,Lihu2012,Brida2010,Jaffrey2008,Bina2013} utilizes the point-to-point intensity fluctuation correlation to retrieve the object's image from the intensity sequence recorded by the single-pixel signal detector. The key idea is the point-to-point intensity fluctuation correlation, which indicates the one-to-one mapping relation. Therefore, intensity correlation can be perfectly used to match feature points. Moreover, for any a point of the left eye image in BPSI, its matching point in the right eye image can be found by intensity correlation.

This article was arranged as follows. Detailed experimental setup and experimental results were demonstrated in Section 2.  Corresponding imaging performance analysis was given in Section 3. And the conclusion was in Section 4.

\section{The experiment}
Different to traditional BPSI, the proposed scheme applies active imaging system. The experimental schematic is shown in Fig.1.
\begin{figure}[htp1]
\centering
\includegraphics[width=100mm, bb=0 0 500 370]{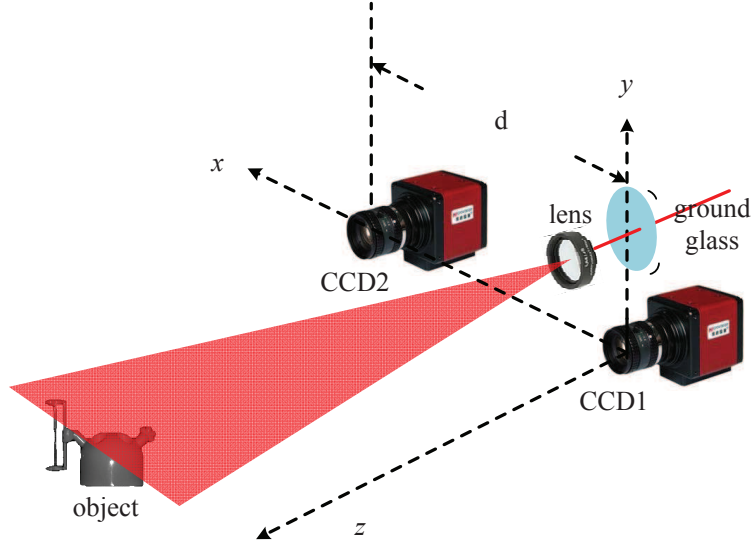}
\caption{\label{scheme} The experimental schematic of BPSI based on correlation matching.}
\end{figure}

The active source used in the experiment was time-varying speckle fields, or pseudo-thermal light, generated by introducing a slowly rotating ground glass after the laser. The wavelength $\lambda$ of the laser ( continuous wave helium-neon laser ) is 633 $nm$. And the radius of the laser beam is about 1$mm$. The ground glass was driven by a step motor with a rotational angular velocity of about 30 degree per second. Located after the ground glass, a collecting lens was used to adjust the divergence angle of the emergent light. Then the object was illuminated by the generated speckle fields.

The capturing devices were the same to traditional BPSI system. Two spatially resolving devices ( CCD ) with the same parameters were used as binocular vision to capture the images of the object. The CCD's pixel number is 512$\times$512, the pixel size is 11$\mu m$, and the field of view is 42.9 degree. CCD1 was located at the left side of the light source, and CCD2 was located at the right side of the light source. These two CCDs were located in parallel along $x$-axis, and their horizontal distance was 145$mm$. According to camera calibration, CCD2 was 120$\mu m$ higher than CCD1 along $y$-axis.

During the imaging procedure, these two CCDs captured pictures synchronously with an exposure time of 8$ms$. For the $i$-th photographing, the captured images of CCD1 and CCD2 were recorded as $LI_i$ and $RI_i$, respectively.

In the experiment, the object was a plastic toy, whose shape is a robot holding a sandglass. The toy was placed in a dark environment and illuminated by the generated speckle fields on the front side. The distance between the toy and the cameras was more than one meter. For a single photographing, the images of the toy captured by CCD1 ( left vision ) and CCD2 ( right vision ) are shown in Fig.2. The parallax between these two images are obvious. According to binocular parallax imaging theory, if the positions of an object point in both the left view and the right view are found, the 3D coordinate of this object point can be calculated. Clearly, the depth resolution along $z$-axis at an object point depends on the precision of matching the positions of this point in disparity maps.
\begin{figure}[htp1]
\centering
\includegraphics[width=100mm]{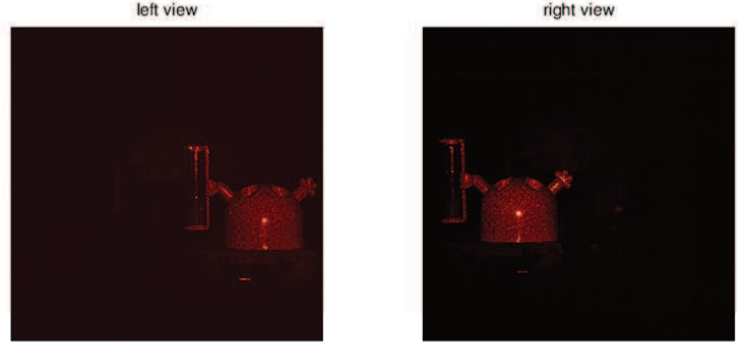}
\caption{\label{binocular} The images of the object in the left and right vision.}
\end{figure}

Because of synchronous sampling, the light intensity distributions on the surface of the object in images $LI_i$ and $RI_i$ are similar speckle pattern. Since the speckle field randomly varies along with the rotation of the ground glass, the speckle patterns are statistically independent. Therefore, for any an object point ($x, y$) in the left-eye views, there is only a correlated point in the right-eye views. And the position of the correlated point in the right-eye views can be found by calculating mutual correlation function, using the intensity values of the point in the left-eye views $LI_i(x, y)$ and the intensity matrixes in the right-eye views $RI_i$,
\begin{equation}\label{CI}
G(x',y')=\langle \Delta LI_i(x,y) \Delta RI_i \rangle = \langle (LI_i(x,y)-\langle LI_i(x,y) \rangle) ( RI_i(x',y') - \langle RI_i(x',y') \rangle ) \rangle, \\
\end{equation}
where $\langle f \rangle=\frac{1}{N}\sum_{i=1}^{N}(f)$, and $N$ is the total photographing number.

The total photographing number $N$ was 1000 in the experiment, and the total exposure time was about 5 second.
By superposing all the images captured by CCD1 ( or CCD2 ), the integral photograph in the left ( or right ) view can be obtained. The integral photograph in the left view is shown in Fig.3. Note that there is an indicator in Fig.3 to point out the position of an object point ( 300, 390 ).
\begin{figure}[htp1]
\centering
\includegraphics[width=100mm]{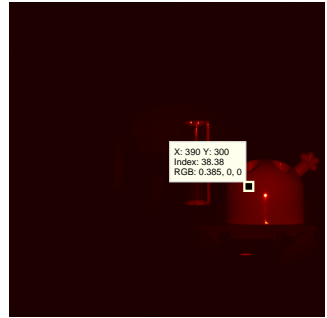}
\caption{\label{integralimage} The integral image of the object in the left view.}
\end{figure}
Apparently, affected by the light source illumination, the brightness distribution of the toy is not uniform. Especially, the brightness at the reflection point is much larger than that at other points. Then in the correlation matching procedure, the reflection point might be misjudged as the matched point. Consequently the 3D coordinate of the object point would be wrong. To avoid this problem, the uniformly weighed correlation imaging algorithm \cite{Lihu2013-2} was applied to calculate the mutual correlation function
\begin{equation}\label{UWCI}
G(x',y')=\langle \Delta LI_i(x,y) \Delta RI_i \rangle = \langle ( LI_i(x,y)-\langle LI_i(x,y) \rangle ) ( \frac{ RI_i(x',y')} {\langle RI_i(x',y') \rangle} - 1 ) \rangle. \\
\end{equation}

Then take the reference point ( 300, 390 ) in the left view for instance, the relative position of the reference point in the left view is shown in Fig.3, and the intensity distribution obtained by calculating Eq.(2) is shown in Fig.4(a). Apparently, the brightest point in Fig.4(a) is the matched point, indicating the position of the object point in the right view. The coordinate of the matched point was found to be ( 288, 124 ). The difference between the line numbers of the reference point and the matched point is 12 pixels, which exactly equals to the height difference between CCD1 and CCD2.
Select a horizontal line across the matched point in Fig.4(a), then the brightness curve of this line is depicted in Fig.4(b). Apparently, the correlation value at the matched point is much larger than that at other points. So the total photographing number as well as the total exposure time can be further decreased in the proposed BPSI system. Moreover, the matching precision equals to the full width at half maximum ( FWHM ) of the bright spot surrounding the matched point, i.e. the correlation length in the object plane. Therefore, the matching precision can easily reach 1 pixel by adjusting the pseudo-thermal light source.
\begin{figure}[htp1]
\centering
\includegraphics[width=120mm, bb=70 285 585 520]{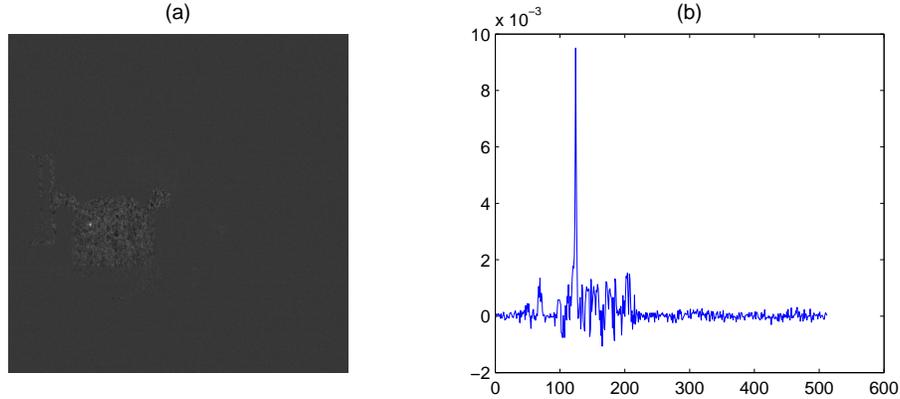}
\caption{\label{correlationimages} (a) The correlated image obtained by using the reference point ( 300, 390 ); (b) the brightness curve of the line across the matched point.}
\end{figure}

Finally, through changing the reference point coordinate in the left view and achieving corresponding matched point's coordinate in the right view, the 3D coordinates of all points on the object's surface of illumination can be calculated. Then the 3D scatter diagrams of the toy obtained in the experiment are shown in Fig.5, in which Fig.5(a) is the side view of the image along $z$-axis, Fig.5(b) is the top view of the image along $-y$-axis, Fig.5(c) is the side view of the image along $x$-axis, and Fig.5(d) is the top view of the image with certain angle of inclination. Fig.5(b) and Fig.5(c) clearly indicate that the depth resolution in the experiment is about 5 $mm$. Besides, from the Fig.5, we can conclude that the depth range of the toy is about from 1165 $mm$ to 1230 $mm$ away from the CCD plane, the thickness of the toy along $z$-axis is about 65 $mm$, the width of the toy along $x$-axis is about 120 $mm$, and the height of the toy along $y$-axis is about 105 $mm$. All these parameters of the toy measured by the proposed BPSI system have little difference with the real values.
\begin{figure}[htp1]
\centering
\includegraphics[width=120mm]{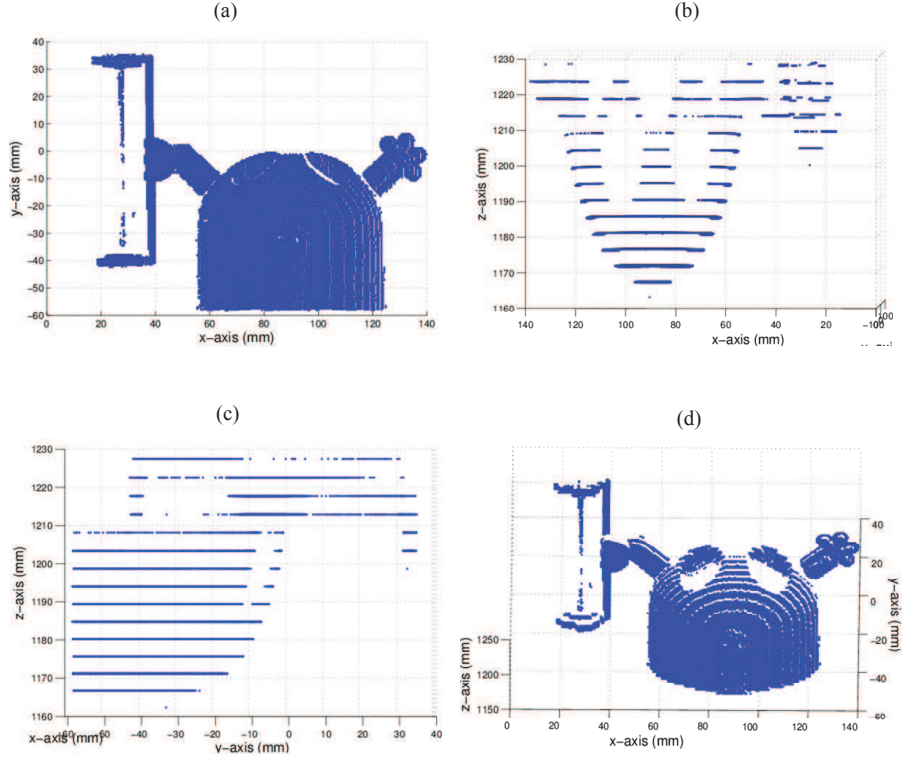}
\caption{\label{scattering} (a)The side view of the image along $z$-axis; (b)the top view of the image along -$y$-axis; (c)the side view of the image along $x$-axis; (d)the top view of the image from certain angle of inclination.}
\end{figure}

\section{Resolution analysis}
In the proposed BPSI system, the transverse resolution along $x$-axis or $y$-axis equals to the transverse correlation length of the speckle field in the object plane. As it is known \cite{Lihu2012,Jaffrey2008,Ferri2005,Han2006,Han2007}, the transverse correlation length at the plane with an axial distance of $z_s$ from the source plane is
\begin{equation}\label{lc}
l_c \approx \frac{\lambda z_s}{D}, \\
\end{equation}
where $D$ is the source size or the beam diameter of the laser illuminated on the ground glass, and $\lambda$ is the wavelength of the laser. And the minimum distinguishable transverse distance at the object plane is
\begin{equation}\label{dx}
dx \approx tan(\frac{\theta}{N_p})z_l, \\
\end{equation}
where $\theta$ is the field of view of the imaging lens, $N_p \times N_p$ is the CCD's pixel number, $z_l$ is the distance between the object plane and the imaging lens plane.

Consequently, as long as $dx > l_c$, the practical transverse resolution of the imaging system will be 1 pixel. Assume that $z_s = z_l$, then the condition is $tan(\frac{\theta}{N_p}) > \frac{\lambda}{D}$. In this experiment,  $tan(\frac{\theta}{N_p}) \approx 1.46\times 10^{-3}$ and $\frac{\lambda}{D} \approx 0.316\times10^{-3}$, so the actually achieved transverse resolution is 1 pixel.

As Fig.1 shows, the back focus of CCD1's camera lens was set as the original point of the 3D coordinate system. Since the field of view of the camera lens is $\theta$, the size of a single pixel is $p$, the total pixel size of the CCD is $L=N_p*p$, then the distance between CCD's detecting plane and the front focus $z_0=\frac{L}{2tan(\theta/2)}$. Assume the 3D coordinate of the back focus of CCD2's camera lens is (d, 0, 0), and the object plane is $z=z_1$.
Then the actual transverse resolution in the object plane satisfies
\begin{equation}\label{deltax}
\delta x=\delta y=\frac{z_1}{z_0}p = \frac{2z_1 tan(\frac{\theta}{2})}{N_p}, \\
\end{equation}
which indicates that the transverse resolution in the object plane only depends on $\theta$, $z_1$ and $N_p$.

Furthermore, for any an object point ($x_1,y_1,z_1$), the depth resolution $\delta z$ has to satisfy
\begin{equation}
(\frac{z_0}{z_1}-\frac{z_0}{z_1+\delta z}) max \{ |d-x_1|,|x_1|,|y_1| \} \approx p, \\
\end{equation}
which means the image point in the left or right view moves a pixel length $p$ when the object point moves to ($x_1,y_1,z_1+\delta z$). Then after simplification, the approximate depth resolution can be calculated as
\begin{equation}
\delta z \approx \frac{pz_1^2}{z_0 max \{ |d-x_1|,|x_1|,|y_1| \}-pz_1}. \\
\end{equation}
Consequently, the transverse resolution and the approximate depth resolution for any an object point can be calculated with system parameters.

Normally, the depth resolution (longitudinal resolution) along $z$-axis is much larger than the corresponding transverse resolution, so improving the depth resolution is more important in BPSI system.

\section{Conclusion}
In conclusion, a correlation matching idea was proposed to be used in BPSI system. With the help of the correlation matching algorithm, every point on the object's surface of illumination is a feature point, and its 3D coordinate can be calculated based on geometrical optics. Moreover, the transverse resolution and the matching precision can reach 1 pixel easily. Therefore compared with traditional BPSI system, the proposed BPSI system performs better in obtaining the 3D information of a real object. Especially when the feature of the image is not obvious, such as that the object has a smooth 3D surface, the traditional method might fail in obtaining the 3D information of the object's surface. Although the computational complexity of correlation matching algorithm used in the experiment is larger than traditional matching algorithms, it can be reduced sharply by increasing the computer memory and utilizing parallel computation. Note that traditional method performs better than the proposed method on the margin areas and corner points, therefore combining these two methods might obtain more integrated 3D information of a real object.

\section*{Acknowledgments}
This work is supported by the Natural Science Foundation of China (Grant No:61471239) and the Hi-Tech Research and Development Program of China (Grant No: 2013AA122901).


\begin{thebibliography}{99}

\bibitem{Burckhardt1968}
C. B. Burckhardt, ``Optimum parameters and resolution limitation of integral photography,'' J. Opt. Soc. Am. A {\bf 58}, 71¨C74 (1968).

\bibitem{Yang1988}
L. Yang, M. McCormick, and N. Davies, ``Discussion of the optics of a new 3-D imaging system,'' Appl. Opt. {\bf 27}, 4529¨C4534
(1988).

\bibitem{Son2005}
J.-Y. Son and B. Javidi, ``Three-dimensional imaging systems based on multiview images,'' J. Display Technol., {\bf 1}(1), 125 -140 (2005).

\bibitem{Xiao2013}
X. Xiao, B. Javidi, M. Martinez-Corral, and A. Stern, ``Advances in three-dimensional integral imaging: sensing, display, and applications,'' Applied Optics, {\bf 52}(4), 546-560(2013).

\bibitem{Javidi2009}
B. Javidi, F. Okano, and J. Y. Son, ``Three-Dimensional Imaging, Visualization, and Display,'' Springer, (2009).

\bibitem{Rubio2011}
J. L. Rubio-Guivernau, V. Gurchenkov, M. A. Luengo-Oroz, et al. ``Wavelet-based image fusion in multi-view three-dimensional microscopy,'' Bioinformatics, {\bf 28}(2), 238-245(2011).

\bibitem{Sung-Keun2013}
Sung-Keun Lee, Sung-In Hong, Yong-Soo Kim, Hong-Gi Lim, Na-Young Jo, and Jae-Hyeung Park, ``Hologram synthesis of three-dimensional real objects using portable integral imaging camera'', Optics Express, {\bf 21}(20), 23662-23670(2013).

\bibitem{Sun2013}
B. Sun, M. P. Edgar, R. Bowman, L. E. Vittert, S. Welsh, A. Bowman, and M. J. Padgett, ``3D Computational Imaging with
Single-Pixel Detectors'', Science, {\bf 340}, 844-847(2013).

\bibitem{Pittman1995}
T. B. Pittman, Y. H. Shih, D. V. Strekalov, and A. V. Sergienko, ``Optical imaging by means of two-photon quantum entanglement,''
 Phys. Rev. A {\bf 52}, R3429-R3432 (1995).

\bibitem{Boyd2002}
R. S. Bennink, S. J. Bentley, and R. W. Boyd, ``¡°Two-Photon¡± Coincidence Imaging with a Classical Source,''
 Phys. Rev. Lett. {\bf 89}, 113601 (2002).

\bibitem{Gatti2004}
A. Gatti, E. Brambilla, M. Bache, and L. A. Lugiato, ``Ghost Imaging with Thermal Light Comparing Entanglement and Classical Correlation,'' Phys. Rev. Lett. {\bf 93}, 093602 (2004).

\bibitem{Wu2009}
X. H. Chen, Q. Liu, K. H. Luo, and L. A. Wu, ``Lensless ghost imaging with true thermal light,'' Opt. Lett. {\bf 34}, 695-697 (2009).

\bibitem{Lihu2012}
Hu Li, Zhipeng Chen, Jin Xiong, and Guihua Zeng, ``Periodic diffraction correlation imaging without a beam-splitter'', Opts. Express {\bf 20}, 2956-2966 (2012).

\bibitem{Brida2010}
G. Brida, M. Genovese and I. R. Berchera, `` Experimental realization of sub-shot-noise quantum imaging", Nature Photonics {\bf 4}, 227-230 (2010).

\bibitem{Jaffrey2008}
J. H. Shapiro, ``Computational ghost imaging,'' Phys. Rev. A {\bf 78}, 061802(R) (2008).

\bibitem{Bina2013}
M. Bina, D. Magatti, M. Molteni, A. Gatti, L. A. Lugiato, and F. Ferri, ``Backscattering Differential Ghost Imaging in Turbid Media'', Phys. Rev. Lett. {\bf 110}, 083901 (2013).

\bibitem{Lihu2013-2}
Hu Li, Jianhong Shi, and Guihua Zeng, ``Ghost imaging with non-uniform thermal light fields'', J. Opt. Soc. Am. A, {\bf 30}(9), 1854-1861 (2013).

\bibitem{Ferri2005}
F. Ferri, D. Magatti, A. Gatti, M. Bache, E. Brambilla, and L. A. Lugiato, ``High-Resolution Ghost Image and Ghost Diffraction Experiments with Thermal Light'', Phys. Rev. Lett., {\bf 94}, 183602(2005).

\bibitem{Han2006}
J. Cheng, S.S. Han, and Y.J. Yan, ``Resolution and noise in ghost imaging with classical thermal light'', Chin. Phys. Soc., {\bf 15}(9), 2002-2006(2006).

\bibitem{Han2007}
M.H. Zhang, Q. Wei, X. Shen, Y.F. Liu, H.L. Liu, J. Cheng, and S.S. Han, ``Lensless Fourier-transform ghost imaging with classical incoherent light'', Phys. Rev. A, {\bf 75}, 021803(R)(2007).


\end{thebibliography}
\end{document}